\documentclass[aps,prd,twocolumn]{revtex4}
\usepackage{graphicx}
\begin{document}

\title{\bf Primordial nuggets survival and QCD pairing}
\author{G. Lugones$^*$ and J. E. Horvath$^{\dag}$\\
\it Instituto de Astronomia, Geof\'{\i}sica e Ci\^encias
Atmosf\'ericas/Universidade de S\~ao Paulo\\
Rua do Mat\~ao 1226, (05508-900) S\~ao Paulo SP, Brazil\\
Email address: $^*$glugones@astro.iag.usp.br ;
$^{\dag}$foton@astro.iag.usp.br}

\vskip5mm
\begin{abstract}
We revisit the problem of boiling and surface evaporation of quark
nuggets in the cosmological quark-hadron transition with the
explicit consideration of pairing between quarks in a color-flavor
locked (CFL) state. Assuming that primordial quark nuggets are
actually formed, we analyze the consequences of pairing on the
rates of boiling and surface evaporation in order to determine
whether they could have survived with substantial mass. We find a
substantial quenching of the evaporation + boiling processes,
which suggests the survival of primordial nuggets for the
currently considered range of the pairing gap $\Delta$. Boiling is
shown to depend on the competition of an increased stability
window and the suppression of the rate, and is not likely to
dominate the destruction of the nuggets. If surface evaporation
dominates, the fate of the nuggets depend on the features of the
initial mass spectrum of the nuggets, their evaporation rate, and
the value of the pairing gap, as shown and discussed in the text.
\end{abstract}

\maketitle

%---------------------------------------------------------------%
\section{Introduction}
%---------------------------------------------------------------%

Approximately $10^{-5}$ seconds after the Big Bang the early
universe was filled with a hot and expanding mixture of elementary
particles. The Universe was composed mainly by photons, charged
leptons, neutrinos, quarks and gluons (and the corresponding
antiparticles) coexisting in thermal and chemical equilibrium
through electroweak interactions. As the Universe expanded, this
mixture cooled down to a critical temperature at which the plasma
of free quarks and gluons converted into hadrons. Early studies of
this transition started in the 1980's \cite{Olive1981,Suh82,Hog83}
and gave a broad-brush picture of the physics involved  (for a
more complete reference list see \cite{Wit84,
DeK84,AH85,KK86,FMA88,BonPan93}).

A very important question is whether the transition is actually
first order, second order or just a crossover. Dramatic effects
are expected in the first case, while a second order or crossover
would be much less spectacular. Lattice numerical simulation is
the best approach currently available for the study of QCD near
the finite temperature transition point. While it has longly been
known that the transition is first order in the case of pure
gluonic calculations (corresponding to infinitely heavy quarks),
and in the case of four light quarks, the actual physical case is
elusive. At present, there are well established non-perturbative
lattice techniques to study this transition at $\mu = 0$ and
$T\neq 0$. The order of the transition is known as a function of
the quark masses showing that the physical point is probably in
the crossover region unless the $s$ quark mass is small (in which
case it should be first order). For recent reviews see
\cite{Kan,Fod} and references therein.

Interesting baryon fluctuations would have been produced by a
first order transition. The two phases need to coexist long enough
for baryon transport to shuffle the baryon number across the phase
boundary. As pointed out in early studies, the onset of the
supposedly first-order transition requires some degree of
supercooling \cite{Hog83}. If the transition is not first order,
no supercooling could possibly occur (even if the equation of
state gave rise to a very rapid change in the energy density) due
to the extremely slow expansion of the Universe.

The generation of primordial isothermal baryon number
inhomogeneities can be  understood within a scenario of cosmic
separation of phases \cite{Wit84,FMA88}. When the universe cools
to the critical temperature $T_{QCD}$ nucleation of bubbles of the
hadron phase could begin. However, it is a general feature of the
nucleation theory that the nucleation probability is not large
enough at the critical temperature but for temperatures below it.
Therefore, the universe supercools below $T_{QCD}$ being still in
the quark phase until the nucleation rate becomes sufficiently
large. After a brief stage of nucleation during which the hadron
bubbles grow and reheat the universe back to $T_{QCD}$, nucleation
is again inhibited due to its low probability and the expansion of
the universe makes hadron bubbles to grow slowly at expenses of
the quark phase. Once hadron bubbles occupy roughly half of the
total volume they are able to collide and merge leaving the
universe with shrinking droplets of quark-gluon plasma immersed in
a hadron matter medium.

The fate of these baryon number inhomogeneities depend on how heat
and baryon number are transported across the transition front
\cite{Wit84,FMA88}. Latent heat (or entropy) could be carried out
by neutrinos, surface evaporation of hadrons (mostly pions) and by
the motion of the transition front which converts volume of one
vacuum into another. The baryon number transport across the
conversion front depends on the bulk properties of both phases and
on the penetrability of the interface (which quantifies the chance
of baryon number to pass from one side of the boundary to the
other).  Estimations of baryon number penetrability have been made
within the frame of the chromoelectric flux tube model
\cite{FMA88,JF,Sumi90,HindPonj}. If the baryon penetrability
indeed happens to be small, it may be possible to accumulate
almost all the baryon number density in the quark-gluon phase (see
below). In such a case, and depending on the parameters, the
inhomogeneities may be large enough to produce strange quark
matter (SQM). This results in a universe in thermal equilibrium
but with an inhomogeneous baryon distribution (i.e. out of
chemical equilibrium).

The study of the extreme case in which quark nuggets form has been
undertaken by a number of authors
\cite{AF85,Mad86,AO89,Sumi91,Mad91,MadOle91,Mad93,ML}. An absolute
upper limit to the baryon number contained in the nuggets is
determined by the size of the cosmological horizon evaluated at
the critical temperature. The simplest estimate yields the
well-known value

\begin{equation}
A_{max} \, = \, 10^{49} \bigg(\frac{100 \, {\rm
MeV}}{T_{QCD}}\bigg)^{2}.
\end{equation}

Actually, the details of the dynamics will determine an initial
mass function at the end of the transition. This is a quite
complicated problem and has not been solved in detail, although
Bhattacharyya et al. \cite{Hind03} presented a series of
calculations showing that the maximum baryon number of the nuggets
is $\sim 10^{43}$ for $T_{QCD} = 150 \, MeV$, which fit
comfortably within the horizon size.

After the QCD phase transition, the temperature in the primordial
Universe is still high enough to allow for evaporation of hadrons
from the surface of the nuggets, and in principle to allow for
boiling (nucleation of hadronic bubbles inside its volume). This
is a consequence of the presence of the $-TS$ term in the free
energy, which disfavors the strange quark matter phase at
intermediate temperatures. It is only at low temperatures ($T
\approx 2 $ MeV) that nuggets begin to be preferred to free
hadrons. Previous work found that boiling is not possible for
reasonable values of the bag constant $B$ since the timescale is
too short for bubble nucleation to take place
\cite{MadOle91,Mad93}. If so, surface evaporation seems to be the
only mechanism able to destroy the primordial nuggets, although
the very survival of these entities may be considered as still
subject to uncertainties. Since it is likely that quarks inside
the nuggets settle in paired states at a relatively high
temperature (see Fig. 1 for a qualitative sketch), we shall
examine in the remaining of this work the effects of quark pairing
on the evaporation/boiling at intermediates temperatures, thus
revisiting the question of nugget survival.

\begin{figure}
\includegraphics[angle=-90,width=9cm,clip]{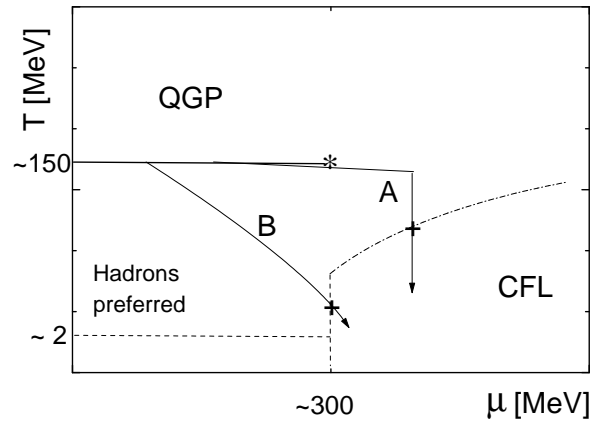}
\caption{Path of the strange quark matter nuggets in the $T - \mu$
plane. Nuggets are quickly formed starting at $T = T_{QCD} \sim
150 $ MeV, and they are fragile to evaporation/boiling at
intermediate temperatures as discussed in text. A transition to
the CFL phase occurs at the points marked with crosses. The path
labelled as "A" assumes a very quick formation of the nugget (that
is, $t_{formation} \ll H^{-1}$, see Ref. \cite{ML}), which evolves
at constant density afterwards following a vertical path. A
(perhaps more realistic) path "B" has been also sketched, in which
the formation is slower $t_{formation} \sim H^{-1}$. After
crossing the CFL boundary nuggets are "safe" because the pairing
now protects them against evaporation/boiling and attaining the
dashed line temperature is no longer relevant for their fate.
Thus, their masses freeze at a higher value when quarks become
locked in CFL states. }
\end{figure}

%---------------------------------------------------------------%
\section{Boiling of CFL nuggets}
%---------------------------------------------------------------%

As stated above, quark nuggets are born hot and therefore
nucleation of hadronic bubbles could occur inside them.
Nevertheless, as it has been shown previously \cite{Mad93},
boiling is unlikely to destroy primordial nuggets for reasonable
values of the bag constant $B$. However, we must note that pairing
must occur at temperatures below the critical temperature
$T_{\Delta} = 0.57 \Delta$. Therefore, the analysis made in
\cite{Mad93} holds only in the temperature regime between
$T_{QCD}$ and $T_{\Delta}$ while below $T_{\Delta}$ pairing
effects must be taken into account. A remarkable consequence of
QCD pairing is that the stability window for strange matter is
considerably enlarged, allowing a wider range for $B$ \cite{LH02}.
Thus, although pairing should difficult boiling because more
energy is necessary to produce an hadron lump, it is not clear
{\it a priori} to what extent the modification of the stability
characteristics of SQM counteracts this effect.

Let us briefly examine the thermodynamic description of boiling of
primordial nuggets including the effect of pairing between quarks.
We assume, for simplicity, that the nucleated phase is in thermal
and chemical equilibrium with the nugget, which itself evolves at
fixed $\mu$ (that is, along a path of the type ``A" in Fig. 1).
The work done to form a bubble of radius $r$ composed by the
hadronic phase inside the quark phase is

\begin{equation}
W = - \frac{4}{3} \pi r^3 \Delta P  + 4 \pi  \sigma  r^2 - 2 \pi
\gamma r + \frac{4}{3} \pi r^3 \times \frac{3}{2} n_B \Delta ,
\end{equation}

\noindent being  $\Delta P = P_h  -  P_q$ the pressure difference
between both phases, $\sigma = \sigma_q + \sigma_h$ the surface
tension, $\gamma = \gamma_q - \gamma_h$ the curvature coefficient,
$n_B$ the baryon number density in the hadronic phase, and
$\Delta$ the gap of the CFL pair. The innovation here is the last
term which is introduced by considering that to put three quarks
in an hadron an energy $\Delta$ must be expended for each CFL pair
that is disassembled inside the quark phase. Further refinements
should be considered if hadrons are assumed to be composed by a
diquark plus a free quark; in this case the net energy released
(per baryon) should be $\sim \frac{1}{2} \Delta$. The effect of
$\Delta$ on $\sigma$ and $\gamma$ themselves is unknown and will
be neglected in this first approach. Note that the gap also enters
the free energy trough the pressure as a quadratic term. The
critical radius $r_c$ is found by extremizing $W$,

\begin{equation}
r_c =  \frac{\sigma}{F} [1 + \sqrt{1 - b} ]
\end{equation}

\noindent being $b \equiv \gamma F / 2 \sigma^2$ and

\begin{equation}
F \equiv \Delta P - \frac{3}{2} n_B \Delta.
\end{equation}

Only those bubbles with a radius greater than $r_c$ will be able
to grow. The nucleation rate for the growing bubbles is given by

\begin{equation}
{\cal R}_{boil} =  T^4 \exp{(-W_c / T)}
\end{equation}

\noindent being $W_c = \frac{4 \pi \sigma^3}{3 F^2}[2 +
2(1-b)^{3/2} - 3b] $. As we shall see below the contribution of
$\Delta$ tends to suppress the rate since it enters in such a way
that $F$ becomes smaller and $W_c$ becomes larger. However, since
the stability behavior is modified by pairing, allowing stability
for a much wider rage of the bag constant $B$, it is necessary to
determine which is the leading effect on the boiling process, as
we shall do in the following.

The effect of boiling on the quark nuggets can be analyzed by
means of a slightly different condition (see \cite{Mad93} and
references therein). Comparing the total area of the nugget and
the bubbles

\begin{equation}
\sum_i A_i^{bubbles} = A^{nugget}
\end{equation}

\noindent it is found that boiling is less important than surface
evaporation for a baryon number $A$ below the value $A_{boil}$
given by

\begin{eqnarray}
A_{boil} & = & 7.90 \times 10^{-61} \bigg( \frac{2 F}{(1 + \sqrt{1 - b} ) T
\sigma} \bigg)^6 \nonumber \\
 & &  \times \exp \bigg( \frac{\pi \sigma^3  [2 + 2(1-b)^{3/2}  -3b]^2  }{T F^2}
\bigg) .
\label{Aboil}
\end{eqnarray}

\noindent For a given value of $A_{boil}$, the last equation gives
the critical bag constant $B$ and $\sigma$ separating the boiling
and the non-boiling regions. The value of the critical $B$ and
$\sigma$ is almost insensitive to the value of $A_{boil}$.
Spanning the range $1 < A_{boil} < 10^{57}$ only changes the
critical values of $B^{1/4}$ and $\sigma^{1/3}$ by a few MeV
\cite{Mad91}.

From an inspection of Eq. (\ref{Aboil}) it is easily recognized
that the main effect of pairing on boiling enters only through a
boost in the bag constant $B$. This can be understood by comparing
the boiling of CFL strange matter with the boiling of unpaired
SQM. If we assume that the strange quark mass $m_s$ is zero, the
pressure of SQM and CFL strange matter differ only by a term
proportional to $\Delta^2$, i.e. $P_{SQM} = P_{CFL} - (3 \Delta^2
\mu^2) / {\pi^2}$ \cite{LH02}. In both, the CFL and the unpaired
SQM phases, there exists a symmetric flavor composition with $n_u
= n_d = n_s$. Note that the last will be not true when considering
that the strange quark has a finite mass; in this case the CFL
phase will still have a symmetric composition but SQM will not.
However, we do not expect strong departures from this simple
analysis. The nucleation of hadron bubbles occurs in a very fast
timescale $\tau_s \sim 10^{-24}$ s, typical of strong
interactions, therefore the transition happens out of weak
equilibrium irrespective of the pairing of quark matter. This
means that flavor will be conserved during the process of
nucleation (only after $\tau_w \sim 10^{-8}$ s, will
$\beta$-decays lead the just formed hadron phase to its
equilibrium configuration). The conservation of flavor during the
formation of bubbles, and the fact of both SQM and CFL strange
matter having the same composition (in a first approximation)
guarantees that exactly the same gas of hadrons will be produced
both beginning with a CFL or with an unpaired SQM composition.
Therefore, the pressure difference $\Delta P_0$ between the SQM
phase and the hadron phase, and the difference $\Delta P$ between
the CFL phase and the hadron one, are related by

\begin{equation}
\Delta P   = \Delta P_0 - \frac{3 \Delta^2 \mu^2}{\pi^2}.
\label{boost}
\end{equation}

\noindent The simple relation given by  Eq. (\ref{boost}) allow us
to gauge straightforwardly the impact of pairing in the boiling
process. The difference $F$ can be written, in the case of
massless quarks, as

\begin{eqnarray}
F & = & \Delta P - \frac{3}{2} n_B \Delta \; \approx \; \Delta P_0
- \frac{3}{\pi^2} \Delta^2 \mu^2 - \frac{3}{2} n_B \Delta \nonumber\\
 & = & P_h - P_{free} - B_{eff}.
\end{eqnarray}

\noindent The effect of pairing has been included in the bag
constant $B$ by defining  an "effective bag constant"

\begin{equation}
B_{eff} = B - \frac{3}{\pi^2} \Delta^2 \mu^2 - \frac{3}{2} n_B
\Delta, \label{B}
\end{equation}

\noindent where $P_{free}$ is the pressure of a flavor-symmetric
mixture of free quarks. Note that pairing enters trough a leading
contribution $\sim \mu^3 \Delta$ associated with the condensation
work and a second order contribution $\sim \Delta^2 \mu^2$
associated with the modification of the pressure in the CFL phase.
Although our analysis does not include the important effects of
the finite mass of the strange quark and of the chemical
composition of the phases, it seems clear that the here shown
tendency should be qualitatively the same in a more realistic
study. Also, as stated above, some refinements would need to be
considered due to the effect of finite temperature, and the
pairing gap in the surface tension and curvature terms. In
summary, all the difference with the boiling of quark nuggets made
up of unpaired massless quarks is that we must use here the
effective bag constant defined by Eq. (\ref{B}).

The likelihood of boiling can be analyzed in the $B$ vs. $T$ plane
(see Fig 2). For an unpaired quark mixture the critical $B$ above
which boiling is allowed (for any baryon number $A$) is always
greater than the maximum $B$ that allows stable SQM for a
transition out of the equilibrium \cite{Mad93}. As we have shown
in Eq. (\ref{B}), the pairing shifts up the critical curve by a
magnitude $({3} \Delta^2 \mu^2) /{\pi^2} +  (3 n_B \Delta) /2$. On
the other hand, as shown in Ref. \cite{LH02}, the maximum $B$ that
allows stable CFL strange matter also shifts up a magnitude $m_n^2
\Delta^2 / (3 \pi^2)$. Therefore, the net shift $h_{\Delta}$ is

\begin{equation}
h_{\Delta} =   \frac{(9 \mu^2 - m_n^2)\Delta^2}{3 \pi^2} +
\frac{3}{2} n_B \Delta,
\end{equation}

\noindent which is clearly positive in the range of interest since
the leading term is the second one and scales as  $\mu^3 \Delta$.
This means that nucleation is impossible in the temperature regime
where QCD pairing operates ($T < T_{\Delta}$), even in the most
favorable situation in which the nucleated phase is in
equilibrium.

%------------------------------------------------------
\begin{figure}
\includegraphics[angle=-90,width=9cm,clip]{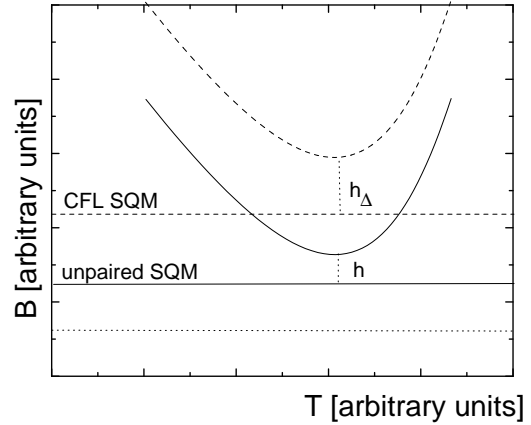}
\caption{Sketch of the effects of CFL pairing on the boiling of
nuggets. The stability window of SQM is realized between the lower
dotted line and the solid horizontal line. Analogously, the wider
stability window of CFL SQM holds between the lower dotted line
and the short-dashed horizontal line as marked \cite{LH02}. For a
given temperature, boiling of primordial quark nuggets is allowed
only above the parabolic-like curves (solid line for unpaired SQM
and dashed for CFL SQM). In spite of the rising of the upper
stability limit for CFL SQM, the boiling curve also raises, and
since both curves lie above the corresponding upper limits for
matter stability (horizontal lines), quark matter nuggets must
survive boiling (with or without pairing).}
\end{figure}

%-------------------------------------------------------------------
\section{Surface evaporation}
%---------------------------------------------------------------%

Surface evaporation of hadrons has been addressed as a first
mechanism for nugget destruction by Alcock and Farhi \cite{AF85}
as explained above. Using simple detailed balance arguments, they
concluded that the flux of baryons from the surface was more than
enough to evaporate the nuggets for all but the highest
(unphysical) masses. Further work revisited the issue by employing
chromoelectric flux tube expressions \cite{Sumi90}, which happened
to be much smaller than the naive flux employed originally. It was
found that nuggets having a baryon number larger than 10$^{39}$
could survive evaporation. Another detailed study by Madsen,
Heiselberg and Riisager \cite{Mad86} also considered explicitly
the effect of flavor non-equilibrium at the surface of the nugget
and found that lumps with  baryon number as low as $A$ $\sim$
10$^{46}$ could survive evaporation. These and other calculations
\cite{JF} suggests that a low surface baryon flux allows the
survival of large, but not extreme nuggets, perhaps down to $A
\sim 10^{40}$ provided the evaporation flux is low enough.
Therefore it is of interest to understand what happens when quark
pairing is introduced in this picture.

An evaporating lump is slightly cooler than the environment,
which for temperatures $\sim 100$ MeV is composed mainly by
photons, neutrinos, electrons and their antiparticles. Heat flows
from the surrounding medium into the lump, providing the energy
to power the evaporation. Near the surface of the lump there is a high
concentration of baryons that have just evaporated. Due to their
large mean free path, neutrinos are the most efficient energy carriers. We
shall discuss the effects of pairing using a scaling of the simplest rate
derived by Alcock and Farhi \cite{AF85}, since the latter provides a
good description of all surface evaporation rates presented in the
literature. The number of hadrons
evaporated from the surface per second is written as

\begin{equation}
{\cal R} =  \alpha \;  \frac{\sigma_0 m_n}{ 2 \pi^2} \; T^{2}
A^{2/3} e^{-I/T} \label{R}
\end{equation}

\noindent with $I=20$ MeV the binding energy, $\sigma_0 = 3.1
\times 10^{-4}$ MeV$^{-2}$ and $m_n$ the neutron mass. The
parameter $\alpha$ is introduced in order to reproduce
approximately the behavior of the flux within very different
models,  which differ by several orders of magnitude (see e.g.
\cite{AF85,Sumi90,HindPonj}).

While above the critical temperature for pair formation
$T_{\Delta}$, the evaporation rate would be given by Eq.(\ref{R}).
Below $T_{\Delta}$, we use the same combinatorial criterion as in
the previous section for the breakup of quark pairs to write down
the rate of evaporation from CFL nuggets ${\cal R}_{\Delta}$ in a
first approximation as

\begin{equation}
{{\cal R}_{\Delta}}= {\cal R} \, \times \, e^{-3\Delta/2T}.
\label{del}
\end{equation}

Since the energy cost of pair breakup is a general feature of the
models, we expect this suppression to hold quite independently of
the detailed physics. The important feature is that surface
evaporation rates get effectively quenched as soon as the CFL
phase sets in, at a temperature $T_{\Delta} = 0.57 \Delta$, which
is certainly much higher than the $\sim \, 2$ MeV necessary to
stabilize the nuggets. Therefore, it may be said that CFL states
freeze out the mass of the nuggets once they cool down to
$T_{\Delta}$.

%%%%%%%%%%%%%%%%%%%%%%%%%%%%%%%%%%%%%%%%%%%%%%%%%%%%%%%%
\begin{figure}
\includegraphics[angle=-90,width=9cm,clip]{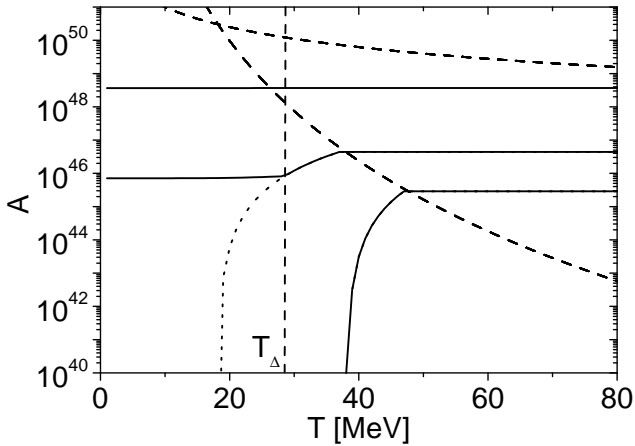}
\caption{The baryon number evolution of quark nuggets as a
function of the temperature of the universe for $\Delta = 50$ MeV.
The upper dashed line is the baryon number included within the
horizon. The lower dashed line (given by $G_F^2 T^4 A^{1/3}
\approx 1$) divides the regions of diffusive neutrino heating and
neutrino transparency of the environment of the nuggets. There is
almost no evaporation in the diffusive neutrino heating regime,
simply because not enough energy is supplied to power the
evaporation. The vertical line shows the temperature $T_{\Delta}$
below which pairing operates. The solid lines show the evolution
of nuggets with three different initial baryon numbers. The
heavier nugget reaches $T_{\Delta}$ even before leaving the
diffusive neutrino heating regime, and therefore freezes out
retaining its initial baryon number. The lighter of these nuggets
evaporates completely as soon as it enters the neutrino
transparent regime. The intermediate mass nugget maintains its
initial baryon number until it enters the neutrino transparent
regime. Thereafter, it evaporates substantially until it reaches
the critical temperature $T_{\Delta}$ where it freezes out with a
smaller mass. This nugget would have evaporated in the absence of
pairing, as is apparent from the dotted curve corresponding to
$\Delta = 0$. These calculations assume  $\alpha = 10^{-3}$, which
is much higher than the values given by chromoelectric flux tube
models but smaller than the extreme value given by detailed
balance.}
\end{figure}

\begin{figure}
\includegraphics[angle=-90,width=9cm,clip]{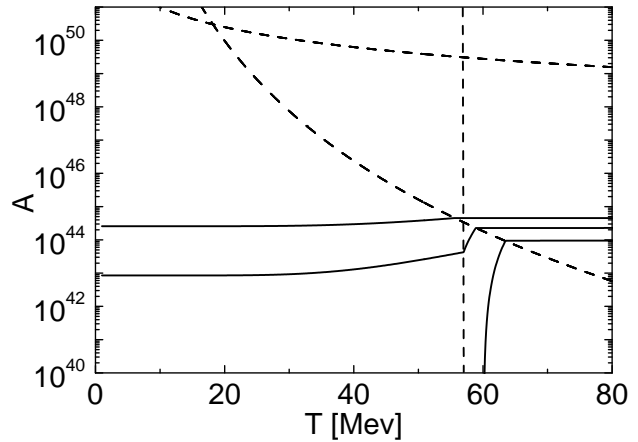}
\caption{The same as the previous figure but for  $\Delta = 100$
MeV. Lower masses can now reach the pairing temperature and freeze
out}
\end{figure}
%%%%%%%%%%%%%%%%%%%%%%%%%%%%%%%%%%%%%%%%%%%%%%%%%%%%

The parametrization of the surface evaporation rates given in Eqs.
(\ref{R}) and (\ref{del}) allows a simple analytic solution of the
evolution equation $dA/dt = {\cal R}$ of the baryon number of the
nugget. This solution is

\begin{eqnarray}
A^{1/3}(T) =  A_0^{1/3}  - C \int_{T}^{T_0} \frac{e^{-I/T}}{T} \;
dT .
\end{eqnarray}

\noindent where we have identified the initial baryon number
$A_{0} \equiv A(T_0)$ and $C = \alpha  (-2 \sigma_0 m_n)/( 6
\pi^2) \times [45 / (172 \pi^3 G)]^{1/2}$ (with $G$ the Newton
constant).

As discussed in Refs. \cite{AF85,Sumi91} the beginning of the
evaporation is possible when the nuggets are surrounded by an
optically thin environment. The neutrino influx is then capable of
powering the baryon evaporation at the nugget surface. The
transition from the diffusive to transparent regime satisfies the
condition $G_F^2 T^4 A^{1/3} \approx 1$ (with $G_F$ the Fermi
constant). Evaporation is small before the nugget crosses this
boundary curve. Thus, the temperature $T_0$ at which each nugget
begin to evaporate is a function of the initial baryon content
$A_0$, assumed to be the value it had at the formation temperature
(see Figs. 3 and 4).

It can be checked that, depending on the value of the pairing gap
$\Delta$, a subset of nuggets that proceed directly from the
diffusive neutrino heating regime to the CFL paired state may
exist. Therefore, these nuggets remain essentially frozen with the
same baryon number they had at their formation. The minimum baryon
number $A_{freeze}$ that satisfies this condition  is found
inserting the relation $T_{\Delta} = 0.57 \Delta$ in the
transition condition above, and is given by

\begin{equation}
A_{freeze} = 3.38 \times 10^{44} \, \bigg( \frac{100 \,{\rm
MeV}}{\Delta}\bigg)^{12}.
\end{equation}

Nuggets smaller than $A_{freeze}$ will evaporate substantially
once they enter the optically thin neutrino regime, but may
survive if they manage to cool down to $T_{\Delta}$ with some
finite mass.

While quarks remain unpaired, the evaporation rate will be  given
by Eq. ({\ref{R}). Therefore, for $T_{\Delta} < T <  T_{0}$ the
baryon number density as a function of temperature is given by

\begin{equation}
A^{1/3}(T) =  A_{0}^{1/3} - C \; [ Ei(-I/T) - Ei(-I/T_{0})]
\end{equation}

\noindent where $Ei(x)$ is the exponential integral.

Once  $T < T_{\Delta}$, pairing reduces the evaporation rate
to the expression Eq.(\ref{del}), and the baryon number density as a
function of temperature follows

\begin{eqnarray}
A^{1/3}(T) & = &  A_{0}^{1/3} \nonumber \\
 & - &  C \; \bigg[ Ei\bigg(\frac{-I - \frac{3}{2} \Delta}{T}\bigg)
- Ei\bigg(\frac{-I - \frac{3}{2} \Delta}{T_{\Delta}}\bigg)\nonumber \\
 & + &   Ei(-I/T_{\Delta}) - Ei(-I/T_{0}) \bigg].
\end{eqnarray}

A simple approximation for $Ei(x)$ which is good within a few
percent in the range of interest, is the following

\begin{equation}
Ei(x) = \frac{\exp(x)}{x^2 - 2 x}
\end{equation}

\noindent which is useful for understanding the relative weight of
each term in the corresponding temperature regimes. The complete
results are depicted in Figs. 3 and 4. The effects of CFL pairing
are apparent when nuggets reach $T_{\Delta}$, since many of them
are able to survive while they would have been evaporated in the
absence of this pairing. As a corollary, we may state quite
generally that a given initial mass function of nuggets would be
stretched towards the smallest masses because of CFL pairing.
Detailed calculations of this features will be presented in a
future publication. We finally stress that this evaporating
population may not exist at all, depending on the form of the
initial mass function.

%-------------------------------------------------------------------
\section{Conclusions}
%---------------------------------------------------------------%

We have discussed in this work the effects of QCD pairing on the
evaporation/boiling rates of quark nuggets assumed to be formed
during the cosmological quark-hadron phase transition. These
nuggets would be produced at $T_{QCD} \sim 150$ MeV with maximum
baryon numbers $A_{max}\sim 10^{49} ({100 \, {\rm
MeV}}/{T_{QCD}})^{2}$ corresponding to the horizon scale at that
epoch. After formation, the nuggets are fragile because of the hot
environment and may boil and/or evaporate into hadrons. The
nuggets may survive if their destruction is not complete when the
Universe cools down to a sufficiently low temperature.

We have shown in this work that the consideration of pairing
brings an additional twist to the problem of nugget survival at
intermediate temperatures. Specifically, we have shown that both
the boiling and the surface evaporation get suppressed because of
the presence of the gap $\Delta$ in the respective rates.

Boiling of nuggets has been already discussed in the literature
and found unlikely in the most realistic calculations. When CFL
pairing is included, the boiling is also unlikely because, in
spite of the increase of the stability window, the rate is
suppressed by $\Delta$ and the net effect produces $h_{\Delta}
> 0$ in realistic cases.

In the case of surface evaporation, the fate of the nuggets depend
mainly on the (unknown) characteristics of the initial mass
spectrum of the nuggets, their evaporation rate, and the value of
the pairing gap. However, and independently of these
uncertainties, many general trends can be noticed. If the value of
the pairing gap $\Delta$ is sufficiently high, the nuggets perhaps
as small as $\sim 10^{42}$ and up to $A_{max}$ enter the CFL phase
before leaving the regime which is opaque to neutrino transport.
Since pairing quenches the rate by a large factor, all these
nuggets freeze out with essentially the same baryon number they
had at formation. In general, the net result is that many nuggets
survive with smaller masses, which could have not otherwise
survived if paring had not operated. Therefore, any initial mass
function of nuggets will be {\it stretched} towards the low-mass
region after being partially evaporated. Note that this behavior
is obtained for evaporating fluxes that may be many orders of
magnitude larger than the very small values indicated by the
chromoelectric flux tube models.

We conclude that the survival of the nuggets (if formed) is much
likely if they settle in a CFL state at a temperature $T_{\Delta}
= 57 \times (\Delta/100 \, {\rm MeV})$ MeV, which may be true for
the whole population. Thus, CFL prevents further
evaporation/boiling and effectively freezes out the masses of the
nuggets. A detailed numerical study of the whole evolution of the
nuggets is desirable to address this issue.

%---------------------------------------------------------------%
\section{Acknowledgements}
%---------------------------------------------------------------%

The authors wish to thank the Instituto de Astronomia, Geof\'\i
sica e Ci\^encias Atmosf\'ericas de S\~ao Paulo, the S\~ao Paulo
State Agency FAPESP for financial support through grants and
fellowships, and the partial support of the CNPq (Brazil).

\newpage


\begin{thebibliography}{99}

\bibitem{Olive1981} K. A. Olive, Nucl. Phys. B 190, 483 (1981)

\bibitem{Suh82} E. Suhonen, (1982) Phys. Lett. B 119, 81 (1982)

\bibitem{Hog83}  C. J. Hogan, Phys. Lett. B 133, 172 (1983)

\bibitem{Wit84} E. Witten, Phys. Rev. D 30, 272  (1984)

\bibitem{DeK84} M. Gyulassy,  K. Kajantie,  H. Kurki-Suonio and L. McLerran,
Nucl. Phys. B 237, 477 (1984)

\bibitem{AH85} J. H. Applegate and C. J. Hogan, Phys. Rev D 31, 3037 (1985)

\bibitem{KK86} K. Kajantie and H. Kurki-Suonio, Phys. Rev. D 34, 1719 (1986)

\bibitem{FMA88}  G. M. Fuller, G. J. Mathews and C. R. Alcock, Phys. Rev. D
37, 1380 (1988)

\bibitem{BonPan93} S. A. Bonometto and O. Pantano, Phys. Rep. 228, 175
(1993).

\bibitem{Fod}  Z. Fodor and  S. D. Katz, hep-lat/0106002 and
hep-lat/0110102 (2001)

\bibitem{Kan}  K. Kanaya, Nucl. Phys. A 715,  233 (2003); Z. Fodor, {\it
ibid.} 715, 319 (2003)

\bibitem{JF} K. Jedamzik and G.M. Fuller, Nuc. Phys. B 441, 215 (1995)

\bibitem{Sumi90} K. Sumiyoshi, T. Kusaka, T. Kamio, and T. Kajino, Phys.
Lett. B 225 (1989) 10 ; K. Sumiyoshi, T. Kajino, G. J. Mathews and
C. R. Alcock, Phys. Rev. D 42, 3963 (1990) .

\bibitem{HindPonj} B. K. Patra, V. J. Menon and C. P. Singh, Nucl. Phys. B 564, 145 (2000)

\bibitem{AF85} C. Alcock and E. Farhi, Phys. Rev. D 32, 1273 (1985)

\bibitem{Mad86} J. Madsen and H. Heiselberg and K. Riisager, Phys. Rev. D  34, 2947 (1986)

\bibitem{AO89} C. Alcock and A. Olinto, Phys. Rev. D  39, 1233 (1989)

\bibitem{Sumi91} K. Sumiyoshi and T. Kajino , in {\it Proceedings of the International
Workshop on Strange Quark Matter in Physics and Astrophysics},
eds. J. Madsen and P. Haensel, {\it Nucl. Phys. B. Proc.
Supp.}{\bf 24}, 80 (1991)

\bibitem{Mad91} J. Madsen, {\it ibid.} \cite{Sumi91} p. 85

\bibitem{MadOle91} J. Madsen and M. Olesen, Phys. Rev. D 43, 1069 (1991)

\bibitem{Mad93} M. Olesen and J. Madsen, Phys. Rev. D  47, 2313 (1993)

\bibitem{ML} L. Masperi and M. Orsaria, astro-ph/0307347 (2003)

\bibitem{Hind03} A. Bhattacharyya, Jan-e Alam, S. Sarkar, P. Roy, B. Sinha, S. Raha
and Pijushpani Bhattacharjee, Phys. Rev. D 61, 083509 (2000)

\bibitem{LH02} G. Lugones and J. Horvath, Phys. Rev. D 66, 074017 (2002)

\end{thebibliography}
\end{document}